\documentclass[twocolumn,showpacs,aps,prl,superscriptaddress,floatfix]{revtex4}
\usepackage{graphicx}% Include figure files
\usepackage{dcolumn}% Align table columns on decimal point
\usepackage{bm}% bold math

\begin{document}

\preprint{}

\title{Spin Waves and Quantum Criticality in the Frustrated XY Pyrochlore Antiferromagnet Er$_2$Ti$_2$O$_7$} 

\newcommand{\eto}{Er$_{2}$Ti$_{2}$O$_{7}$}
%\draft 
\author{J.P.C. Ruff}
\affiliation{Department of Physics and Astronomy, McMaster University,
Hamilton, Ontario, L8S 4M1, Canada}
\author{J.P. Clancy} 
\affiliation{Department of Physics and Astronomy, McMaster University, 
Hamilton, Ontario, L8S 4M1, Canada}
\author{A. Bourque}
\affiliation{Department of Chemistry, Institute for Research in Materials,
  Dalhousie University, Halifax, Nova Scotia, B3H 4J3, Canada}
\author{M.A. White}
\affiliation{Department of Chemistry, Institute for Research in Materials,
  Dalhousie University, Halifax, Nova Scotia, B3H 4J3, Canada}
\author{M. Ramazanoglu}
\affiliation{Department of Physics and Astronomy, McMaster University,
Hamilton, Ontario, L8S 4M1, Canada}
\author{J.S. Gardner} 
\affiliation{National Institute of Standards and Technology, 
100 Bureau Drive, MS 6102, Gaithersburg, MD 20899-6102.}
\affiliation{Indiana University, 2401 Milo B. Sampson Lane, Bloomington, Indiana 47408, USA}
\author{Y. Qiu}
\affiliation{National Institute of Standards and Technology, 
100 Bureau Drive, MS 6102, Gaithersburg, MD 20899-6102.}
\affiliation{Department of Materials Science and Engineering, University of Maryland, College Park, Maryland 20742, USA}
\author{J.R.D. Copley}
\affiliation{National Institute of Standards and Technology, 
100 Bureau Drive, MS 6102, Gaithersburg, MD 20899-6102.}
\author{H.A. Dabkowska}
\affiliation{Department of Physics and Astronomy, McMaster University,
Hamilton, Ontario, L8S 4M1, Canada}
\author{B.D. Gaulin}
\affiliation{Department of Physics and Astronomy, McMaster University,
Hamilton, Ontario, L8S 4M1, Canada} 
\affiliation{Canadian Institute for Advanced Research, 180 Dundas St. W.,
Toronto, Ontario, M5G 1Z8, Canada}

\begin{abstract} % insert abstract here 

We report detailed measurements of the low temperature magnetic phase 
diagram of Er$_2$Ti$_2$O$_7$.  Heat capacity and time-of-flight neutron scattering 
studies of single crystals, subject to magnetic fields applied along the 
crystallographic [110] direction, reveal unconventional low energy states.  
Er$^{3+}$ magnetic ions reside on a pyrochlore lattice in Er$_2$Ti$_2$O$_7$, where local XY anisotropy and antiferromagnetic interactions give rise to a unique frustrated system.  In zero field, the ground state exhibits coexisting short and long range order, accompanied by soft collective spin excitations previously believed to be absent.  The application of finite magnetic fields tunes the ground state continuously through a landscape of non-collinear phases, divided by a zero temperature phase transition at $\mu_0 H_c \sim$ 1.5 T.  The characteristic energy scale for spin fluctuations is seen to vanish at the critical point, as expected for a second order quantum phase transition driven by quantum fluctuations.

\end{abstract} 

\maketitle 
% \narrowtext
% \twocolumn
% body of paper here

%intro stuff
Despite intense study spanning more than a decade, highly frustrated magnetism in the rare-earth-metal titanate pyrochlores defies a thorough understanding\cite{greedan}.  The pyrochlore lattice is a network of corner-sharing tetrahedra, and the archetype for geometrical frustration in three dimensions\cite{diep}.  The rare-earth titanates in particular have a tendency towards novel physics, due to a conspiracy of energy scales where dipolar interactions and crystal field anisotropy are of comparable importance to the exchange energy between magnetic moments.  Exotic behaviours attributed to these pyrochlores include spin ice\cite{GBrev} and spin liquid\cite{Gardner99} ground states, magnetoelastic fluctuations near zero temperature\cite{ruffxray}, emergent magnetic monopole excitations\cite{monopole}, and order-by-disorder\cite{champion,gingrasXY}. 

\begin{figure} 
\centering 
\includegraphics[width=8.5cm]{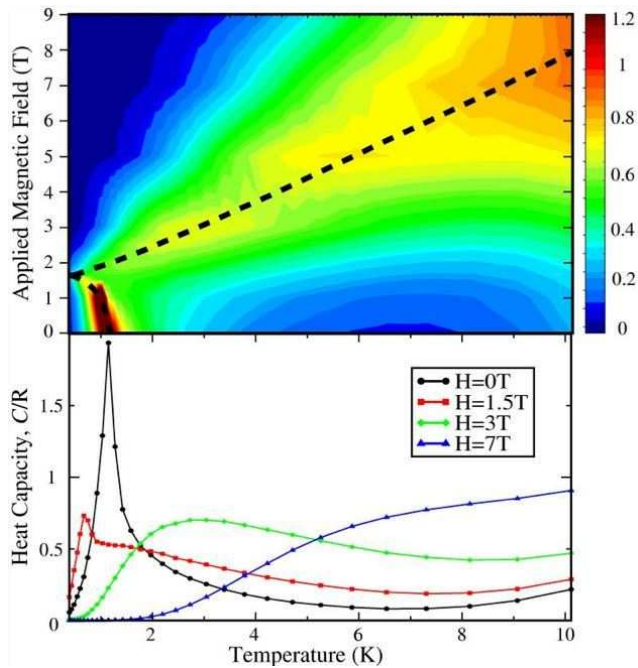}
\caption{The phase diagram of {\eto} in an applied [110] magnetic field is 
shown.  The contour plot was constructed from multiple heat capacity curves, on both warming and cooling.  A sampling of these is shown in the bottom panel.  Dotted lines on the top panel trace out the location of the maximum at each field magnitude.  It is only when $H < H_c$ that this boundary line represents a true phase transition.}
\label{Figure 1}
\end{figure}

Previous studies of {\eto} identify an antiferromagnetic insulator, with a 
Curie-Weiss temperature of $\theta_{CW} \sim$ -22 K\cite{champion,blote}.  
Crystal field analysis yields XY anisotropy such that the magnetic moments 
of the Er$^{3+}$ ions, with effective magnitude $\sim 
3\mu_B$\cite{champion}, are confined to planes tangential to the local tetrahedra on which they reside.  Heat capacity\cite{blote} and neutron diffraction\cite{champion,wills} measurements on polycrystalline samples show a continuous phase transition to long range order at $T_c \sim 1.1$ K in zero field.  Using the group theoretical language of previous work\cite{champion,wills}, the observed magnetic order is a six-fold degenerate ground state labeled under the basis vector $\psi_2$ of the irreducible representation $\Gamma_5$.  This ground state is predicted for a simple nearest neighbour exchange model with planar anisotropy\cite{champion}, where it has been shown to be selected by thermal fluctuations in an example of what Villain described as order-by-disorder\cite{villain}.  However, it is surprising that this simple model should accurately describe {\eto}, where dipolar interactions are expected to be an important perturbation and should select a different ground state\cite{palmerchalker}.  Therefore, the possibility that this ground state is selected energetically by a more complicated microscopic Hamiltonian, rather than via order-by-disorder, has not been ruled out.

Single crystals of {\eto} were grown using the floating zone technique and 
a two-mirror image furnace, in 3 atm. of air and at a growth rate of 5 mm h$^{-1}$.  Heat capacities were measured for a 3 mg crystal using the relaxation calorimetry method with a $^3$He insert.  A similar 7.3 g single crystal was selected for the neutron measurements, which were performed using the NG4 Disk Chopper Spectrometer\cite{DCS} at the National Institute of Standards and Technology Center for Neutron Research.

Heat capacity measurements of single crystal {\eto} (Fig. 1) in zero 
applied magnetic field recreate the previously reported anomaly at $T_c \sim$ 1.1 K, which signals the second order phase transition to long range magnetic order.  Under the application of modest magnetic fields, the anomaly remains sharp, but shifts to lower temperature, finally disappearing at a critical field value $\mu_0 H_c$ slightly larger than 1.5 T.  For fields of 1.75 T and above a broad feature resembling a Schottky anomaly is seen, at a temperature that depends linearly on the applied field strength.  This is suggestive of a crossover to a \textit{quantum paramagnetic} high-field state, reminiscent of that of the prototype transverse field quantum critical magnet LiHoF$_4$\cite{sachdev,aeppli_1,aeppli_2}.

Knowing the location of the phase boundaries, we now focus on the nature 
of the different phases.  To this end, high quality single crystals of 
{\eto} were studied by the time-of-flight neutron scattering technique.  This allows a simultaneous measurement of both the elastic and inelastic neutron scattering cross-section within the [H,H,L] plane in reciprocal space.  Elastic scattering reveals the magnetic structure of the ordered phases, while inelastic scattering informs on the elementary spin excitations of these states.  Pre-existing neutron scattering measurements\cite{champion,wills} identified the zero-applied-field ordering wavevector as the (2,2,0) Bragg reflection.  This result is confirmed in Fig. 2, where a dramatic enhancement of scattering intensity at (2,2,0) is observed upon cooling below the temperature of the heat capacity anomaly.  However, we also note that a second, broad component to the scattering is evident.  As is clear from Fig. 2 f),  when subjected to an applied magnetic field, the magnetic reflection becomes sharper and the background decreases substantially.  This is unambiguous evidence for the existence of a second length scale in the zero-field ground state, where short-range magnetic correlations coexist with long-range magnetic order.  The width of this broad feature corresponds to spin correlations of 15 Angstroms ($\sim$ 1-2 unit cells). The origin of these short-range correlations is not understood, but their presence at T=50 mK is consistent with strong quantum fluctuations in the ground state.  Muon spin rotation results support this, as they observe an anomalously large relaxation rate below T$_c$\cite{musr}.

\begin{figure} 
\centering 
\includegraphics[width=8.5cm]{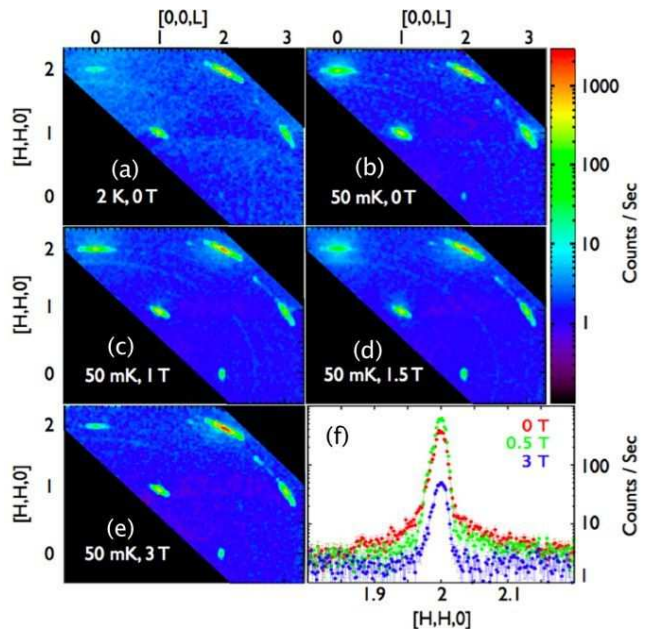}
\caption{Elastic scattering is shown within the [H,H,L] plane for panels a) - e). Note that only nuclear allowed Bragg positions show intensity throughout the phase diagram.  Panel f)  shows longitudinal cuts through (220) at 50 mK.  The existence of a broad, diffuse component to the scattering in zero applied magnetic field is clear, indicating short range magnetic correlations that coexist with long range order.}
\label{Figure 2}
\end{figure}

Within the [H,H,L] plane, we have measured the magnetic scattering intensity of five distinct Bragg peaks at low temperature (T = 50 mK) as a function of applied magnetic field (Fig. 3c).  In weak fields, the scattering is dominated by the large (2,2,0) peak associated with the zero magnetic field ground state (Fig 3a).  An anomalous increase is observed in the scattering intensity at the ordering wavevector for $\mu_0 H =$ 0.5 T.  This effect has been ascribed to the creation of a single domain state in previous work\cite{champion}, however we attribute it to the elimination of the short-range correlations that we have observed in zero field (see Fig. 2).  As we approach $H_c$ from below, the intensity at (2,2,0) drops off smoothly as long range order is destroyed.  Simultaneously, new intensity develops at the (2,2,2) position, which dominates the scattering in high magnetic field.  Strong scattering at (2,2,2) is consistent with a highly polarized XY state, as illustrated in Fig. 3b.  The boundary between the (2,2,0) and (2,2,2) dominated phases coincides with the quantum critical point deduced from heat capacity in Fig. 1.  However, the field-induced reorganization of scattering intensity is not sharp at the critical point.  Rather, a smooth evolution is seen for fields within $\sim$ 1 Tesla of $\mu_0 H_c$.  We also observe an increase in scattering intensity at the (0,0,2) position for $H \sim H_c$, although this magnetic Bragg peak is weak at both low and high fields.

\begin{figure} 
\centering 
\includegraphics[width=8.5cm]{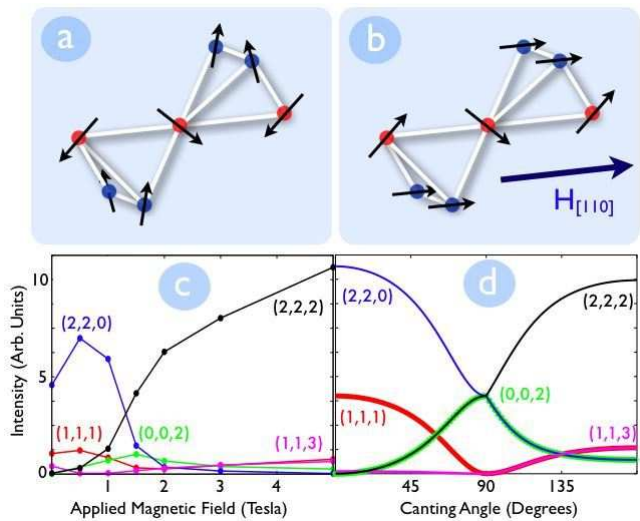}
\caption{(a) Schematic diagram of one of six degenerate ground states in zero field. (b) A schematic diagram of the high-field ground state.  The spins are fully polarized within XY planes and the state is non-degenerate.  Note that spins are subdivided into two sets, residing on perpendicular chains and coloured red and blue.  (c) The measured field dependence of the integrated magnetic Bragg intensity of Q=0 peaks at T=50 mK.  The nuclear component to the scattering is eliminated by subtracting data taken at 2K.  (d) Calculated magnetic Bragg scattering is shown for a smooth deformation of the magnetic structure taking it from a) to b).  A canting angle of 0$^\circ$ corresponds to a) while 180$^\circ$ corresponds to b).}
\label{Figure 3}
\end{figure}

The field evolution of the magnetic scattering is qualitatively understood as a smooth deformation of the ground state from configuration A (illustrated in Fig. 3a) to configuration B (illustrated in Fig. 3b).  To visualize this, it is useful to subdivide the pyrochlore lattice into spins residing on two sets of orthogonal chains, coloured red and blue in Fig. 3 a) and b).  To go from A to B, it is necessary to rotate all blue spins by 90$^\circ$ such that they polarize along $\vec{H}$, while half of the red spins must rotate by 180$^\circ$ and half need not move at all.  A simple picture emerges in which all spins initially rotate through 90$^\circ$, fully polarizing the blue spins while maintaining antiferromagnetic correlations between the red spins.  This leads to a polarized coplanar state, where blue spins point along [1,1,0] and red spins point along $\pm$ [1,-1,0].  Next, while the blue spins remain pinned along the [1,1,0] direction, the red spins rotate through another 90$^\circ$, maximizing their projections along the applied field direction and smoothly increasing the scattering at (2,2,2).  This picture gives rise to the calculated Bragg intensities displayed in Fig. 3d.
  
The above scenario describes the evolution of only one of the six degenerate low-field states into the unique polarized high field state.  A more realistic picture needs to take into account coexisting domains of all six A-type states subject to a global symmetry-lowering [1,1,0] magnetic field.  In this context, the quantum critical point at $H_c$ is most likely related to a field-induced merging of ground states, as recently predicted for Gd-based pyrochlores\cite{ramirez}.  Nevertheless, we expect the qualitative description of a manifold of smoothly field-deformed states to survive more sophisticated analysis, given the striking similarities between the observed and calculated Bragg intensities.

\begin{figure} 
\centering 
\includegraphics[width=8.5cm]{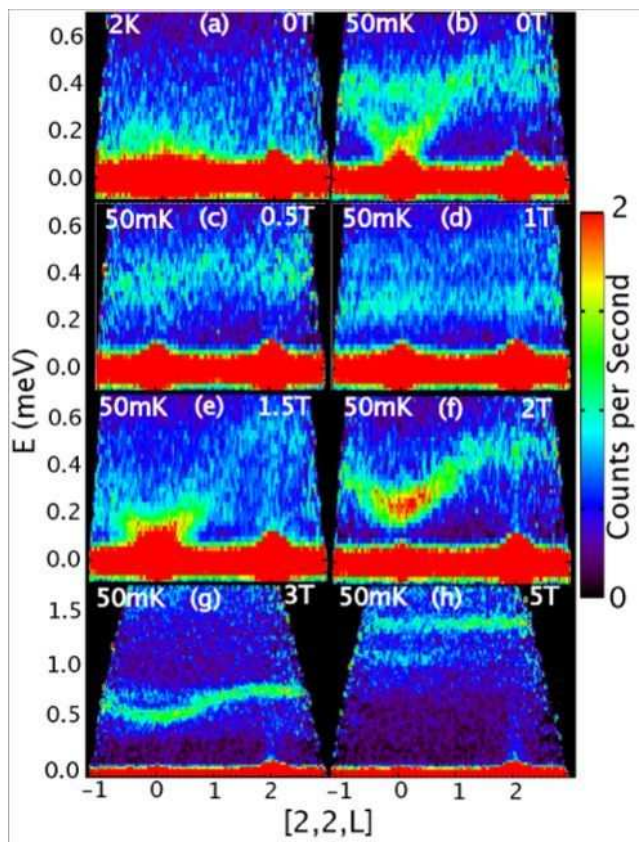}
\caption{The dispersion of spin waves is shown along the [2,2,L] line in 
reciprocal space, joining the (220) and (222) wavevectors, where Bragg scattering characteristic of the low and high field low temperature states is observed.  a) shows data at T=2 K and H=0, while b) - h) show data at T=50 mK and applied field as indicated.  A full spin excitation softening is observed for $\mu_0 H \sim$ 1.5 T.}
\label{Figure 4}
\end{figure}

Complementary to our diffraction results, we report inelastic neutron 
scattering measurements of the low-energy excitations of {\eto} across the 
phase diagram.  Excitations have been measured throughout the [H,H,L] 
plane, and a subset of these results are reported in Fig. 4.  In zero 
applied magnetic field, the onset of magnetic order is accompanied by the appearance of well defined spin wave excitations.  Two modes are easily seen near the (2,2,0) position, one being essentially dispersionless at $\sim$ 0.4 meV while the other goes soft at the ordering wavevector, identifying it as a Goldstone mode.  The dispersionless mode was reported previously, along with the \textit{absence} of the soft mode\cite{champion}.  It has been argued that the unusual ground state of {\eto} lacks soft spin excitations, or that the lowest energy modes fail to couple to the neutron\cite{champion}.  Here, we explicitly show that gapless, Goldstone spin excitations exist, since we observe spin waves which go to zero energy at (2,2,0), (1,1,1) and (1,1,3).  This unambiguously resolves a long-standing point of confusion: unconventional as it may be, the zero-field ground state does indeed support soft spin excitations.  

Under the application of finite magnetic fields, the excitation spectrum 
displays a rich evolution.  At $\mu_0 H =$ 0.5 T, the increase in elastic 
scattering at (2,2,0) is accompanied by the elimination of the soft mode at the same position in reciprocal space.  Clearly, the application of such weak fields does not simply create a single domain state, but rather it induces a qualitative change in the ground state.  In this regime, spin excitations are still seen to go soft at (1,1,1) and (1,1,3), but not at (2,2,0).  Application of stronger fields splits the dispersionless feature into two flat modes, the lower of which decreases towards zero energy transfer.  This mode softening is expected at a second order quantum critical point, where the energy gap should vanish as a power law in $\mid H-H_c \mid$\cite{sachdev}.  Interestingly, a full mode softening is avoided in the canonical case of LiHoF$_4$ due to hyperfine coupling which leaves the quantum critical region inaccessible\cite{aeppli_2}.  This does not occur in {\eto}.  At $\mu_0 H = $ 1.5 T, near $H_c$, the system is ungapped to within our instrumental resolution.  An intriguing picture of coexisting phases emerges, where critical fluctuations of the low field state are observed around (2,2,0), accompanied by a precursor of the intense dispersive excitation characteristic of the high field state.  This mode is clear at 2 T and 3 T, showing that the high field ground state does not correspond to a simple disordered paramagnet.  Certainly, the difference between the excitation spectra of the quantum paramagnetic phase (Fig. 4f), and that of the thermal paramagnet (Fig. 4a) is undeniable.  In larger applied fields, the ground state is dominated by Zeeman energy.  This is evident at 5 T, where two dispersionless modes are seen.  The existence of two such modes is indicative of the subdivision of the pyrochlore lattice discussed earlier, where the Zeeman energy of polarized red and blue spins are not equivalent due to local planar anisotropy.

To conclude, the low temperature phase diagram of {\eto} supports a rich 
variety of elementary excitations as the ground state is tuned by external applied field.  The most dramatic variation with occurs around $\mu_0 H_C \sim$ 1.5 T, where a seemingly continuous quantum critical point is observed.  The degree to which this landscape of low energy states and quantum critical behaviour can be understood will hinge upon a proper energetic description of the system, and we hope this work will guide theoretical efforts in that pursuit.

We thank M.J.P Gingras, A.G. del Maestro and K.A Ross for useful discussions.
We acknowledge the technical assistance of E. Fitzgerald and M.B. Johnson, and
the use of the DAVE software package for elements of the data reduction and
analysis \cite{dave}.  Facilities used are supported by the National Science
Foundation under agreement no. DMR-0454672, the Canadian Foundation for
Innovation, and the Atlantic Innovation Fund.  This work was supported by NSERC of Canada.

% now the references. delete or change fake bibitem. delete next three
%   lines and directly read in your .bbl file if you use bibtex.

% that you will use with ref{} command in the braces of the \label{} command.
%

% Uncomment the following to put the figures into the latex file.

%\begin{figure}
%\centering
%\includegraphics[width=8.5cm]{fig4_nolines_tilt.ps}
%\caption{Enter caption here.}
%\label{fig4}
%\end{figure}

%\begin{figure}
%\centering
%\includegraphics[angle=0,origin=c,width=8cm]{all4_2.ps}
%\caption{Caption}  
%\label{fig2}
%\end{figure}

%\begin{figure}
%\centering
%\includegraphics[angle=0,origin=c,width=8cm]{th2thputtogether2.ps}
%\caption{Caption}
%\label{fig3}
%\end{figure}

%\begin{figure}
%\centering
%\includegraphics[angle=0,origin=c,width=8cm]{dspace.ps}
%\caption{Caption}
%\label{fig4}
%\end{figure}

\end{document}